\documentclass{article}

\usepackage{arxiv}

\usepackage[utf8]{inputenc} 
\usepackage[T1]{fontenc}    
\usepackage{hyperref}       
\usepackage{url}            
\usepackage{booktabs}       
\usepackage{amsfonts}       
\usepackage{nicefrac}       
\usepackage{microtype}      
\usepackage{lipsum}

\usepackage[backend=biber,style=numeric,sorting=ynt,maxbibnames=99]{biblatex}
\usepackage{graphicx}
\usepackage{multirow}
\usepackage{wrapfig}
\usepackage{amsmath}

\title{Tools for modelling and simulating the Smart Grid}

\author{
  Ricardo M. Czekster\\
  Newcastle University, School of Computing \\
  \texttt{ricardo.melo-czekster@newcastle.ac.uk} \\
}

\usepackage{longtable}
\usepackage{multicol}
\usepackage{tablefootnote}

\newcommand{\STAB}[1]{\begin{tabular}{@{}c@{}}#1\end{tabular}}
\newcommand{\rotate}[1]{\STAB{\rotatebox[origin=c]{90}{#1}}}

\begin{document}
\maketitle

\begin{abstract}
The Smart Grid (SG) is a Cyber-Physical System (CPS) considered a critical infrastructure divided into cyber (software) and physical (hardware) counterparts that complement each other.
It is responsible for timely power provision wrapped by Information and Communication Technologies (ICT) for handling bi-directional energy flows in electric power grids.
Enacting control and performance over the massive infrastructure of the SG requires convenient analysis methods.
Modelling and simulation (M\&S) is a performance evaluation technique used to study virtually any system by testing designs and artificially creating `what-if' scenarios for system reasoning and advanced analysis. 
M\&S avoids stressing the actual physical infrastructure and systems in production by addressing the problem in a purely computational perspective.
Present work compiles a non-exhaustive list of tools for M\&S of interest when tackling SG capabilities.
Our contribution is to delineate available options for modellers when considering power systems in combination with ICT.
We also show the auxiliary tools and details of most relevant solutions pointing out major features and combinations over the years.
\end{abstract}


\section{Introduction}
Modelling and simulation (M\&S) is a key technique for assessing the performance of virtually any system.
The Smart Grid (SG) is a Cyber-Physical System (CPS) and also a critical infrastructure amenable for M\&S.
The SG encompasses a large infrastructure through extensive use of Information and Communication Technologies (ICT).
Gungor et al. (2012)~\cite{gungor2012survey} surveyed the SG in terms of ICT requirements, explaining its major features and concerns for modelling assets and new service capabilities for increased remote control.
Alongside Industrial Control Systems (ICS) and Supervisory Control and Data Acquisition (SCADA) systems, they pose a large complex system requiring multiple analysis angles~\cite{stouffer2011guide}.
M\&S provides means to artificially create models that map real-world settings where analysts address `what-if' questions and quantitative analysis over multiple scenarios.

Simulation was revamped in past years to accommodate the embedding of power systems with telecommunication, giving rise to coupled simulations, or co-simulations for short.
This technique is used to combine two or more simulation engines into one global simulation as stipulated by Gomes et al. (2018)~\cite{gomes2018co}.
The authors have surveyed the literature and discussed major principles behind this technique.
Co-simulation is used in the SG~\cite{farhangi2009path,lin2011power,fang2011smart,gungor2011smart,le2019smart} to synchronise a continuous time Electric Power Simulator (EPS) with a discrete time telecommunication network simulator.
Co-simulation is a multifaceted mechanism for investigating hybrid systems from distinct application domains, besides only power or telecommunications.
For instance, it may integrate multi-purpose simulation engines, according to the problem under consideration.

Mets et al. (2014)~\cite{mets2014combining} explained the need for co-simulation in power grids whereas Vogt et al. (2018)~\cite{vogt2018survey} discussed available tools and the main mechanism behind hybrid engines whilst tackling the so called `synchronisation problem'.
Specific to energy management, Azeroual et al. (2020)~\cite{azeroual2020simulation} investigated primarily wind turbines in microgrids discussing major tools for modelling multi-agent systems.
Energy assets are spread out over large geographical domains operating as Distributed Energy Resources (DER).
These can be based on Renewable Energy Resources (RER) such as solar roof-top Photo-Voltaic (PV) or Wind Turbines Power Converter Systems (WTPCS), or storage mechanisms in batteries statically installed in buildings, or mobile ones present in Electric Vehicles (EV).

As massive numbers of DER are scattered over the infrastructure, telecommunication becomes a problem when relying on energy assets to regulate frequency as well as other so called ancillary services provision~\cite{greenwood2017frequency}.
One disconcerting problem is when malicious cyber incursions attempt to delay packets traversing the network, corrupt data, or prevent them from reaching their destinations as discussed in a wealth of work throughout the years~\cite{nagaraju2017survey,humayed2017cyber,radoglou2019securing,kumar2019smart}.

The contribution of this work is to point out relevant M\&S tools for EPS, telecommunications, co-simulation engines, and modelling mechanisms for use in the SG.
We highlight the most used ones over the years, discussing their major features and internal details, listing the auxiliary tools and libraries on their design.

The work is divided as follows.
Section~\ref{s:tools} describes modelling and simulation tools for use in the Smart Grid.
In Section~\ref{s:discussion} we present a discussion on tools and models, highlighting advantages and shortcomings.
Section~\ref{s:conclusion} ends our work with our final considerations.

\section{Simulation tools}\label{s:tools}
We describe next proprietary (Section~\ref{ss:prop}) and open source or free co-simulation tools (remaining sections).
Our focus is directed on modelling power and telecommunication systems.
The chosen order to present results are by ascending publication year, i.e., since it was first proposed to the research community.

\subsection{Proprietary}\label{ss:prop}
Private companies and enterprises implemented substantial power based simulation solutions over the years.
A tool is considered proprietary if it used at least one proprietary solution in the mix with other open-source and free libraries or implementations.
These close solutions have obfuscated their tool's internal schema and details.

\begin{itemize}
   \item \textbf{Electric power simulators:} Examples are Siemens' Power System Simulator (PSS)\footnote{PSS\textregistered power system simulation and modeling software. Link: \url{https://new.siemens.com/global/en/products/energy/energy-automation-and-smart-grid/pss-software.html}.} and PowerWorld\footnote{Link: PowerWorld >> The visual approach to electric power systems\url{https://www.powerworld.com/}.}.
Another significant tool is the DiGISILENT PowerFactory\footnote{Link: \url{https://www.digsilent.de/en/powerfactory.html}.} for simulating power systems~\cite{gonzalez2014powerfactory}.
   \item \textbf{Telecommunication networks simulators:} A tool called OPNET Modeler\footnote{OPNET. Link: \url{https://support.riverbed.com/content/support/software/opnet-model/modeler.html}.} was used to help users define and work with different topologies and technologies.
   \item \textbf{Hardware-in-the-loop (HIL) approaches:} 
   \begin{itemize}
       \item \textbf{1995:} A solution worth mentioning is the Real-Time Digital Simulator (RTDS)~\cite{kuffel1995rtds}, a proprietary solution implementing a real-time simulation library for EPS and HIL modelling.
       \item \textbf{2005:} OPAL-RT is a real-time EPS with HIL  capabilities~\cite{dufour2005hardware}\footnote{Power System Simulation Software. Link: \url{https://www.opal-rt.com/power-systems-overview/}.}. It was coupled with the telecommunication network simulator provided by OPNET~\cite{bian2015real}.
   \end{itemize}
   \item \textbf{Co-simulation:}
   \begin{itemize}
       \item \textbf{2006:} In terms of co-simulation, we highlight Electric Power and cOmmunication synCHronizing Simulator (EPOCHS)~\cite{hopkinson2006epochs}. It uses ns-2, Power Systems Computer Aided Design (PSCAD)~\cite{anaya2002modeling} and for transient simulation, EMTDC (electromagnetic) and the PSLF (electromechanical). EPOCHS offers a version with reduced features for free/testing.
       \item \textbf{2012:} Global Event-driven CO-simulation framework (GECO)~\cite{lin2012geco} uses General Electric's Positive Sequence Load Flow (PSLF), Optimal Power Flow (OPF), and ns-2 to tackle SG co-simulation. GECO is not free because it employs proprietary tools in its solution.
       \item \textbf{2013:} The INtegrated co-Simulation of Power and ICT systems for Real-time Evaluation (INSPIRE)~\cite{georg2013inspire} uses other proprietary software such as OPNET Modeler and DIgSILENT PowerFactory. On the same direction, the Testbed for Analyzing Security of SCADA Control Systems (TASSCS)~\cite{mallouhi2011testbed} uses OPNET and PowerWorld for cybersecurity in SCADA.
       \item \textbf{2014:} Another tool is the Secure Operation of Sustainable Power Systems Simulation Platform (SOSPO-SP)~\cite{morais2014sospo}. It assesses security and stability of EPS employing proprietary tools and libraries such as PSS and RTDS for modelling SCADA and Phasor Management Units (PMU). Sun et al. (2014)~\cite{sun2014co} combined OpenDSS and OPNET for studying the reliability of control strategies in the SG composed by a high number of DER.
       \item \textbf{2017:} The European ERIgrid project~\cite{van2017cyber} implemented a tool that used the Functional Mock-up Interface (FMI)/Functional Mock-up Units (FMU) and devised a holistic approach combining PowerFactory, Mosaik, OpenModelica, MATLAB/Simulink, and ns-3 altogether. Saxena et al. (2017)~\cite{saxena2017cpsa} devised a cyber-security tool for framing risk assessments in the SG for better situational awareness. It combined PowerWorld, MATLAB, GridSim, Matlabcontrol, JADE, and ns-3 into its framework.
   \end{itemize}
   \item \textbf{Virtual testbeds:}
   \begin{itemize}
       \item \textbf{2009:} The Virtual Power System Testbed (VPST)~\cite{bergman2009virtual} focused on network security analysis in SCADA. The main idea behind VPST is to provide modellers with a virtual framework for experimenting with equipment reactions in extreme (e.g. under attack) situations in large-scale power and telecommunication networks. For power it used PowerWorld whereas for telecommunications it employed Rinse~\cite{liljenstam2005rinse}, a real-time virtual network to model the cyber part of the electrical grid.
      \item \textbf{2011:} VPNET~\cite{li2011vpnet} uses the Virtual Test Bed (VTB) with OPNET, and presented a case study using a converter model and communication issues, whereas the Multi-Agent and Communication Simulator (MAC-sim)~\cite{perkonigg2013mac} used Java Agent DEvelopment (JADE) platform~\cite{bellifemine2001jade} and OPNET Modeler.
   \end{itemize}
\end{itemize}

\vspace{0.2cm}
Obvious advantages of proprietary software are usually the contractual support and constant maintenance, frequent updates and patches for defects correction, as well as fast and timely execution.
The tools are usually shipped to customers with a plethora of examples and modelling possibilities, and well as a comprehensive documentation.
The tools are carefully validated, so they offer higher accreditation, however, since it is a closed solution, model interoperability over other software suites is limited.
As another disadvantage, we mention that sometimes costs are prohibitive, as the licensing model often entails buying a Commercial-Off-The-Shelf (COTS) solution, even though the customers will only use a fraction of the software.

Next we describe open-source and free solutions available for modellers, dividing the analysis into EPS, telecommunication, general purpose, co-simulators, Transactive Energy Systems (TES), MATLAB, and other tools, as follows.

\subsection{Electric Power Simulators}\label{ss:eps_sim}
\noindent$\bullet~$\textbf{2007:} 
The Internet-technology Based Power System Simulator (InterPSS)~\cite{zhou2007internet,zhou2013distributed,zhou2017interpss} is written in Java used for on-line real-time EPS.
It supports cascading failure analysis, forecasting and CIM\footnote{The Common Information Model (CIM) consists of European and international standards.} model processing.

\noindent$\bullet~$\textbf{2008:} 
GridLAB-D~\cite{chassin2008gridlab,chassin2014gridlab} is an agent based EPS for energy distribution networks implemented at the Pacific Northwest National Laboratory (PNNL) in the US.
It is written in C++, except for some scripting and MATLAB interfaces, which contributes to its performance capabilities.
It is a scalable solution and well documented as modellers have access to several examples and IEEE Test Feeders to use as basis for more complex analysis.

\noindent$\bullet~$\textbf{2010:} 
MATPOWER\footnote{MATPOWER - Free, open-source tools for elec. power syst. simul. and optimization. Link: \url{https://matpower.org/}.} has wide acceptance among industry, enterprises, and academia ~\cite{zimmerman2010matpower}.
It offers MATLAB scripts for steady-state analysis of Power Flow (PF), Continuation Power Flow (CPF), extensible Optimal Power Flow (OPF), Unit Commitment (UC), and stochastic, secure multi-interval OPF/UC.

\noindent - PYPOWER\footnote{PYPOWER - PyPI. Link: \url{https://pypi.org/project/PYPOWER/}.} is a Python solution derived from MATPOWER whereas oct2pypower\footnote{Link: \url{https://github.com/rwl/oct2pypower}.} serves as a Python bridge to MATPOWER.

\noindent$\bullet~$\textbf{2012:} 
The Open Distribution System Simulator (OpenDSS)\footnote{OpenD. Link: \url{https://www.epri.com/pages/sa/opendss}. According to the website, the tool has been used since 1997.} is also a relevant EPS tool~\cite{montenegro2012real}.
It supports modelling DER for grid modernisation and integration.

\noindent$\bullet~$\textbf{2016:} 
Krishnamurthy (2016)~\cite{krishnamurthy2016psst} developed the Power System Simulation Toolbox (psst)\footnote{Link: \url{https://github.com/ames-market/psst}.} to use in conjunction with AMES (see Section~\ref{ss:tes}).
It offered a Graphical User Interface (GUI) for modelling and it was implemented in Python.

\noindent$\bullet~$\textbf{2018:} 
pandapower\footnote{pandapower. Link: \url{http://www.pandapower.org/}.} is also written in Python, promising higher level of modelling automation~\cite{thurner2018pandapower}.
It is shipped with a comprehensive set of examples and a detailed documentation so modellers may create and adapt models to their concerns.

\noindent The Scalable Electric Power System Simulator (SEPSS)~\cite{ilic2018scalable} employed MATLAB in a modular fashion to ease analysis.
It was designed for multi-layered modeling of complex system where it allowed analysts to design multi-temporal cyber signals such as disturbances, market signals, and control set points.

\subsection{Telecommunication networks simulators}\label{ss:telecom}
\noindent$\bullet~$\textbf{2010:} 
The Network Simulator ns-3~\cite{riley2010ns} has high acceptance in both industry and academia.
The latest version has multiple telecommunication modelling capabilities for diverse sets of protocols and packet transmission technologies (wired, wireless, or other medium).
Despite being introduced in 2010, with its constant updates and improvements it is still a major player.

\noindent - Another tool worth mentioning is the Objective Modular Network Testbed in C++ (OMNET++) that together with the INET Framework~\cite{varga2010omnet} adds consistent telecommunication modelling when modelling complex networks\footnote{Link for OMNeT++: \url{https://omnetpp.org}, and for INET: \url{https://inet.omnetpp.org/}.}.

\noindent - The Network Security Simulator (Nessi)~\cite{schmidt2010application} is a network simulator built for security concerns.
It offers plug-ins for profile-based automated attack generation, traffic analysis and support for detection algorithms.
The main difference between other network simulators is to offer a detection Application Programming Interface (API) for integrating and evaluating Intrusion Detection Systems (IDS).

\subsection{General purpose co-simulators}
General Purpose (GP) co-simulators are implementations that could virtually model any real-world problem considering the restrictions imposed by the modelling abstractions.

\noindent$\bullet~$\textbf{2016:} 
The Integrated Tool Chain for Model-based Design of CPS (INTO-CPS)~\cite{larsen2016integrated} was integrated into the co-simulator offered by Maestro~\cite{thule2019maestro} using Modelio Multi-Model, OpenModelica, SysmML, and 20-sim.
It employs the FMI/FMU standard~\cite{blochwitz2011functional} as the middleware for communicating across the toolchain.

\noindent$\bullet~$\textbf{2018:} 
Multi-agent Environment for Complex SYstems CO-simulation (MECSYCO)~\cite{camus2018co}, written in Java (a C++ version is mentioned to be under development), is a GP simulator for hybrid systems modelling.
The tool uses the formal Discrete eVent system Specification (DEVS) for communicating among simulators~\cite{zeigler1989devs,gon2000theory}.

\subsection{Power/telecommunications co-simulation engines}
\noindent$\bullet~$\textbf{2011:} 
Mosaik\footnote{mosaik - A flexible Smart Grid co-simulation platform. Link: \url{https://mosaik.offis.de/}.} is a co-simulation framework that promises to incorporate flexibility for the creation of large-scale scenarios for EPS~\cite{schutte2011mosaik}.
It has bindings to PYPOWER and a demonstration scenario for modellers to learn how to use the tool.

\noindent - The SCADAsim~\cite{queiroz2011scadasim} framework is built on top of OMNeT++ and using MATLAB/Simulink models adapted from SimPowerSystems\footnote{The tool is available at \url{https://github.com/caxqueiroz/scadasim}, however, it has not being updated since 2011.}.

\noindent - The co-simulation proposed by Liberatore et al. (2011)~\cite{liberatore2011smart,al2012comprehensive} used Modelica for the electric grid and ns-2 for modelling the telecommunications.

\noindent - Grunewald et al. (2011) improved Nessi (Section~\ref{ss:telecom}) and developed Nessi2~\cite{grunewald2011agent}, adding improvements over previous versions.
The latter was extended to work with InterPSS~\cite{chinnow2011simulation}, where the authors applied it in cybersecurity.

\noindent$\bullet~$\textbf{2012:} 
The Smart-Grid Common Open Research Emulator (SCORE)~\cite{tan2012score} offered an emulation environment for EPS, with the possibility of integrating power and telecommunication systems.
It provided mechanisms for energy model programming interfaces such as shiftable and non-shiftable loads, and renewable energy resources (PV and WTPCS).

\noindent$\bullet~$\textbf{2014:} 
The Framework for Network Co-Simulation (FNCS)~\cite{ciraci2014fncs,huang2017open} combines GridLAB-D with ns-3 for co-simulating power distribution and telecommunication.
It synchronises simulation engines acting as a firmware to coordinate time passage and interaction.

\noindent - GridSpice~\cite{anderson2014gridspice} is a cloud-based solution for power transmission and generation that allowed analysts to scale designs as the platform executed in a pay-as-you-go model.
It used GridLAB-D with MATPOWER in a graphical interface equipped with a geographical editor, project explorer, and wizard for importing objects from other tools.
The MATPOWER dealed with the transmission and economic dispatch problem whereas GridLAB-D was used for power distribution in a master/worker schema (a supervisor catering several simultaneous tasks).

\noindent - Awad et al. (2014)~\cite{awad2014sgsim} proposed SGsim, a co-simulation framework using OMNeT++ and OpenDSS.

\noindent$\bullet~$\textbf{2015:} 
Bytschkow et al. (2015) combined GridLAB-D, CIM, and AKKA (Java-based co-simulation engine) to model a SCADA system~\cite{bytschkow2015combining}.

\noindent - Hansen et al. (2015) developed Bus.py~\cite{hansen2015bus}, a GridLAB-D communication interface for EPS.
It was used to simulated the communication between a set of customers in a distribution network and an aggregator.
Bus.py also enabled the co-simulation of Energy Management Systems (EMS) and the simulation of Integrated Transmission and Distribution (ITD) systems.

\noindent$\bullet~$\textbf{2016:} 
Zambrano proposed GridTeractions~\cite{zambrano2016gridteractions}, a framework for teaching and testing grid designs using a hardware-software architecture with miniprocessor terminals (Raspberry PI 2).
The solution used OpenDSS through a graphical auxiliary front-end called OpenDSS-G\footnote{A graphical view of OpenDSS. Link:\url{https://sourceforge.net/projects/dssimpc/}.} in a co-simulation with LabVIEW~\cite{montenegro2012real}.

\noindent - Hannon et al. (2016)~\cite{hannon2016dssnet} developed the Distribution System Solver Network (DSSnet) using Mininet's SDN emulation for telecommunications and OpenDSS for EPS.
The authors presented the details behind the architecture and a case study on load shifting.

\noindent$\bullet~$\textbf{2017:} 
FNCS is currently being replaced by a multi-federated approach called the Hierarchical Engine for Large-scale Infrastructure Co-Simulation (HELICS)~\cite{palmintier2017design} by the same research group.

\noindent - Python for Power System Analysis (PyPSA)~\cite{brown2017pypsa,horsch2018pypsa} is used for power flow and electric power system steady-state analysis.
The authors compared their solution with existing (and similar) approaches in terms of grid and economical analysis as well as comparing PyPSA with MATPOWER for several examples.

\noindent$\bullet~$\textbf{2018:} 
The Cyber physical co-simulation platform for Distributed Energy Resources in smart grids (CyDER)~\cite{coignard2018cyder,nouidui2019cyder,gehbauer2019cyder} is a modular co-simulation framework for addressing DER in the SG.
It is under development by a consortium of academic and industrial partners lead by the Lawrence Berkeley National Laboratory (US).
CyDER is based on the FMI standard and GridDyn with some Python integration (PyFMI) for PV and EV modelling, building simulators and PMU.
The technical report does not mention how the telecommunication among DER operates, however, the platform is still under development.
We mention the effort to integrate buildings with co-simulation through FMU as described by Nouidui et al. (2018)~\cite{nouidui2018simulatortofmu}.

\noindent - Hantao and Fangxing (2018)~\cite{cui2018andes,cui2020hybrid} implemented Andes\footnote{Link: \url{https://github.com/cuihantao/andes}.}, a Python-based co-simulation that worked with MAPOWER and PSAT combined with Mininet\footnote{Mininet: An Instant Virtual Network on your Laptop (or other PC). Link: \url{http://mininet.org/}.}, a Software Defined Networks (SDN) communication networks simulator.
It employed PyPMU for M\&S of networks of PMU/PDC in large scale settings.
Andes opened different input models such as PSS/E raw format, MATPOWER, and Dome.

\noindent - Krishnamoorthy and Dubey~\cite{krishnamoorthy2018framework} presented a co-simulation framework combining MATPOWER with OpenDSS for modelling the impact of DER over ITD systems.
The authors validated the approach against DIGiSILENT PowerFactory~\cite{velaga2018transmission}.

\noindent$\bullet~$\textbf{2019:} 
The Zero OBvious Node Link co-simulator (ZerOBNL)~\cite{puerto2019zerobnl}\footnote{Link: \url{https://github.com/IntegrCiTy/zerobnl}.} was implemented in Python and used pandapower in a so called partitioned approach, breaking down elements into more manageable parts in larger co-simulations.
They employed ZeroMQ~\cite{hintjens2013zeromq} for communicating among elements, a fast socket library for distributed applications and Docker for faster virtualised executions.
This framework is part of a larger research project called IntegrCiTy\footnote{Link: \url{https://github.com/IntegrCiTy}.}, offering a large set of examples and auxiliary tools for EPS.

\noindent - Pallonetto et al. (2019)~\cite{pallonetto2019simapi} took a different direction and developed a framework called SimApi for benchmarking building control algorithms.
It employs the Building Controls Virtual Test Bed (BCVTB), a software environment that integrates different simulation engines into a global co-simulation\footnote{Here is a list of software that are linked to BCVTB: EnergyPlus, Modelica and Dymola (simulation environment), FMU/FMI 1.0 and 2.0, MATLAB/Simulink, Radiance ray-tracing software for lighting analysis, ESP-r for building energy modeling, TRNSYS (system simulation program), and the BACnet stack. Link: \url{https://simulationresearch.lbl.gov/bcvtb}.}.
It is cloud-based (i.e. it runs online) and allows for co-simulation of EMS and Building Energy Systems (BES).
There is no comment on the software used for EPS used for executing the platform (it is encapsulated within the system).

\noindent - Steinbrink et al. (2019)~\cite{steinbrink2019cpes} used Mosaik to devise a testing platform for experimenting in Cyber-Physical Energy Systems (CPES).
It allowed modellers to conduct planning, quantify uncertainty and assist domain experts in collaboration efforts.

\noindent - Kennouche et al. (2019) developed the Asynchronous Smart Grid Simulation (ASGriDS)\footnote{Link: \url{https://github.com/taqen/asgrids/tree/appeec19}.}~\cite{kennouche2019asgrids}, a framework that combined pandapower with GNU/Linux based emulators and tools to model asynchronous communication between elements.
It can be deployed with different telecommunication models, i.e., emulated, simulated, or real, where modellers could for example model network impairments with GNU/Linux-based traffic and shaping tool (\texttt{tc})~\cite{hubert2002linux} and the \texttt{netem} module~\cite{hemminger2005network} (for network emulation).

\noindent$\bullet~$\textbf{2020:} 
The SCEPTRE toolchain~\cite{johnson2020assessing} is a virtualised (on-line) option for modelling DER and investigate communication latencies across technologies.

\noindent - The Open Platform for Energy Networks (OPEN)~\cite{morstyn2020open} uses pandapower internally and presents two examples for modellers, one consisting of an EV set up, and another mimicking an EMS.

\noindent - The SmartGrid Cosimulation Platform~\cite{de2020co} uses ns-3 combined with OpenDSS and Mosaik.
It presented two case studies, one for voltage regulation and another for setting up the power grid with telecommunications.

\subsection{Transactive Energy Systems}\label{ss:tes}
Modelling the market within TES is a crucial element in the SG~\cite{greer2014nist} as if offers customers the ability to trade power in the grid as prosumers (both consumers and producers)~\cite{gungor2012survey,huang2018simulation}.
Many authors have considered modelling of TES and its operational details throughout the years, under economical, engineering, and computing scopes.

\noindent$\bullet~$\textbf{2009:} 
Researchers addressed modelling power systems shortcomings by implementing an open-source software solution known as Agent-based Modeling of Electricity Systems (AMES) Wholesale Power Market Test Bed~\cite{li2009development}.
It consists of a framework for non-power specialists unfamiliar with the intricacies of EPS, bringing together economy related aspects into the modelling.
AMES allows working with strategic trading behaviour for addressing engineering and economical problems in the grid.

\noindent$\bullet~$\textbf{2016:} 
Palmintier et al. (2016)~\cite{palmintier2016igms} proposed the Integrated Grid Modeling System (IGMS).
It focused on the automated generation of large-scale scenarios in wholesale market simulations in Transmission-Distribution interactions.
IGMS used the Flexible Energy Scheduling Tool for Integration of Variable Generation (FESTIV)~\cite{ela2012studying} for representing bulk markets, MATPOWER, and GridLAB-D.
The case studies attest the scalability of the system by allowing hundreds of transmission nodes and over a million distribution nodes, where the authors showed examples of varying size IEEE Test Feeders.

\noindent$\bullet~$\textbf{2018:} 
The Transactive Energy Simulation Platform (TESP)\footnote{For more information on the framework and documentation, visit \url{https://github.com/pnnl/tesp/}.}~\cite{huang2018simulation} aggregated GridLAB-D, ns-3 (using FNCS and HELICS) and it also provided means to work with buildings modelled in EnergyPlus~\cite{crawley2001energyplus}.
EnergyPlus is a comprehensive building simulator addressing thermodynamics with wide acceptance and accreditation across stakeholders.
In TESP, all those auxiliary tools work as libraries that are executed and linked together for executing the platform.
It has an extensive list of examples and IEEE Test Feeders to extend, enjoying all the features already available in GridLAB-D as mentioned earlier.

\noindent$\bullet~$\textbf{2019:} 
The ITD-TES Platform~\cite{nguyen2019integrated} addressed modelling the wholesale energy market in the transmission level having one or more distribution systems attached to it as DER (or grid edge resources).
It was used to evaluate designs of transactive energy in ITD systems and it employed FNCS as the co-simulator engine.

\noindent - The ETX~\cite{zhang2019etx} framework worked with energy trading and management implementing communication with \texttt{sockets.io}.
It used trained neural networks for power-flow calculations, aiming at coordinated and repeatable investigations of market, operations, and regulations actions\footnote{Unfortunately, the paper did not explain the internals of the software behind ETX's operation}.

\subsection{MATLAB}
MATLAB (MATrix LABoratory) was developed by MathWorks, and Simulink is its visual interface.
As mentioned in Section~\ref{ss:eps_sim}, MATPOWER is an EPS with large adoption in the research community.
Next, we will mention other relevant MATLAB-based supporting tools for both simulation and co-simulation.

\noindent$\bullet~$\textbf{2008:} 
Power System Analysis Toolbox (PSAT)\footnote{Power System Analysis Toolbox. Link: \url{http://faraday1.ucd.ie/psat.html}.} is a MATLAB toolbox for power system analysis and simulation~\cite{milano2008open}.

\noindent$\bullet~$\textbf{2010:} 
MatDyn\footnote{MatDyn - Electa. Link: \url{https://www.esat.kuleuven.be/electa/teaching/matdyn/}.} is a MATPOWER inspired dynamic analysis free MATLAB script for EPS~\cite{cole2010matdyn}.

\noindent$\bullet~$\textbf{2013:} 
Milano presented Dome~\cite{milano2013python}, based on PSAT (which employed MATLAB), and discussed the suitability of using Python in complex power analysis and studied its performance in a IEEE 14-bus Test Feeder.

\noindent$\bullet~$\textbf{2014:} 
GridMat~\cite{al2014gridmat} offered a visual interface within MATLAB to model GridLAB-D elements for power and control algorithms using Simulink.

\noindent$\bullet~$\textbf{2015:} 
Amarasekara et al. (2015) used MATLAB/Simulink (module SimPowerSystems) and ns-3~\cite{amarasekara2015co} in a co-simulation.
The authors developed a custom mediator (an agent) to coordinate events between Simulink and ns-3 exchange model.
They presented a case study consisting of a small scale microgrid where they studied energy scheduling and the effects of packet losses in communication.

\noindent$\bullet~$\textbf{2016:} 
MATLAB/Simulink and ns-3 was also chosen by Pan et al. (2016)~\cite{pan2016ns3}, where the authors presented an example of a load reconfiguration.
The authors compared results against the GECO framework, discussing how their solution stands out and the reasoning behind modelling choices.

\noindent$\bullet~$\textbf{2017:} 
The integrative Power and Cyber Systems (iPaCS)~\cite{ravikumar2017ipacs} testbed used MATTRANS for power system modelling, a MATLAB/Simulink transient stability program combined with ns-3 (based on MATPOWER)\footnote{MATTRANS - A MATLAB/Simulink Power System Transient Stability Simulation Package. Link: \url{https://github.com/gelliravi/MatTrans}.}.
It studied the latency between Phasor Management Units (PMU) attached to Phasor Data Concentrators (PDC) and generated datasets for deeper analysis.
The authors commented the testbed's suitability for investigating i) latency and bandwidth; ii) congestion between PMU and PDC systems, and; iii) impact of cyber attacks on EPS.

\noindent$\bullet~$\textbf{2018:} 
V{\'e}lez-Rivera et al. (2018)~\cite{velez2018gorilla} proposed Gorilla, an open interface for co-simulation that is agnostic of EPS or telecommunications network simulation tool.
The authors offered a set of simple operations to use such as read/write, call, and subscribe/call-back, for diverting simulation control.
They created two examples, one for voltage regulation, and another for controlling a home microgrid, calling MATLAB/Simulink commands, however, there was no mention on modelling telecommunications.

\subsection{Related tools}
There is a number of auxiliary tools to complement other toolsets/toolkits, as explained next.

\noindent$\bullet~$\textbf{Visual editors for modelling power systems:}
The Open Modelling Framework (OMF)\footnote{Link: \url{https://omf.coop/}.} provides a graphical interface to define huge and complex networks (not bounded by power systems).
The tool is on-line and require registering, helping modellers devise trade-off analysis (for instance, amounts of PV and WT to meet some power demand), or scaled up GridLAB-D models, to mention a few.

\noindent - Webgme-gridlabd\footnote{Link: \url{https://github.com/finger563/webgme-gridlabd}.} offers a visual front-end for creating GridLAB-D models.
The tool employs Node.js and it works on an Internet browser.
It is possible to download the tool and install into a local server.

\noindent$\bullet~$\textbf{Other tools worth mentioning:}
DSATools\footnote{Link: \url{https://www.dsatools.com/}.} is a proprietary tool for a wide range of electric-related assessments.
PYPOWER (no longer maintained) is a power simulator written in Python and used in other tools\footnote{Link: \url{https://pypi.org/project/PYPOWER/}.}.

We also mention the cyber-DEfense Technology Experimental Research (Deter)~\cite{benzel2006experience}, an on-line platform for conducting cybersecurity experiments\footnote{Link: \url{https://deter-project.org/}.}.

\subsection{Co-simulation frameworks for domain specific applications}
Researchers have created bespoke co-simulation frameworks to address domain specific applications such as cybersecurity.
The sheer size of CPS realities mandates its consideration and attack impact analysis in both power and telecommunication networks.

\noindent$\bullet~$\textbf{2016:}
The Attack Simulation Toolset for SG Infrastructures (ASTORIA)~\cite{wermann2016astoria} framework modelled SCADA sub-systems, most notably Master Terminal Units (MTU) and Remote Terminal Units (RTU).
It was developed for Mosaik using PYPOWER and ns-3 focusing on cybersecurity.
ASTORIA used \textit{Attack Profiles} to model adversaries and presented two case studies, one emulating a malicious software infection attack, and another in a Denial-of-Service (DoS) study.

\noindent$\bullet~$\textbf{2020:} 
Cybersecurity concerns in the SG was also addressed in GridAttackSim~\cite{le2020gridattacksim}, which used FNCS to model cyber-attacks.
It offered a visual tool for modelling systems using Python graphical bindings.
As malicious incursions, the authors modelled a channel jamming and a False Data Injection attack (FDI) involving dynamic pricing (energy market).
The tool extended examples provided by FNCS and GridLAB-D where the novelty was to present a GUI for modellers.

\section{Discussion}\label{s:discussion}
Since the inception of the SG, researchers addressed M\&S as a means to reason about operational decisions and test new designs or better configurations.
That is one the reasons as to why the literature on power, telecommunications, and couple tools have witnessed a steep increase over past years.

\begin{figure}[h]
	\centering
	\includegraphics[width=\textwidth]{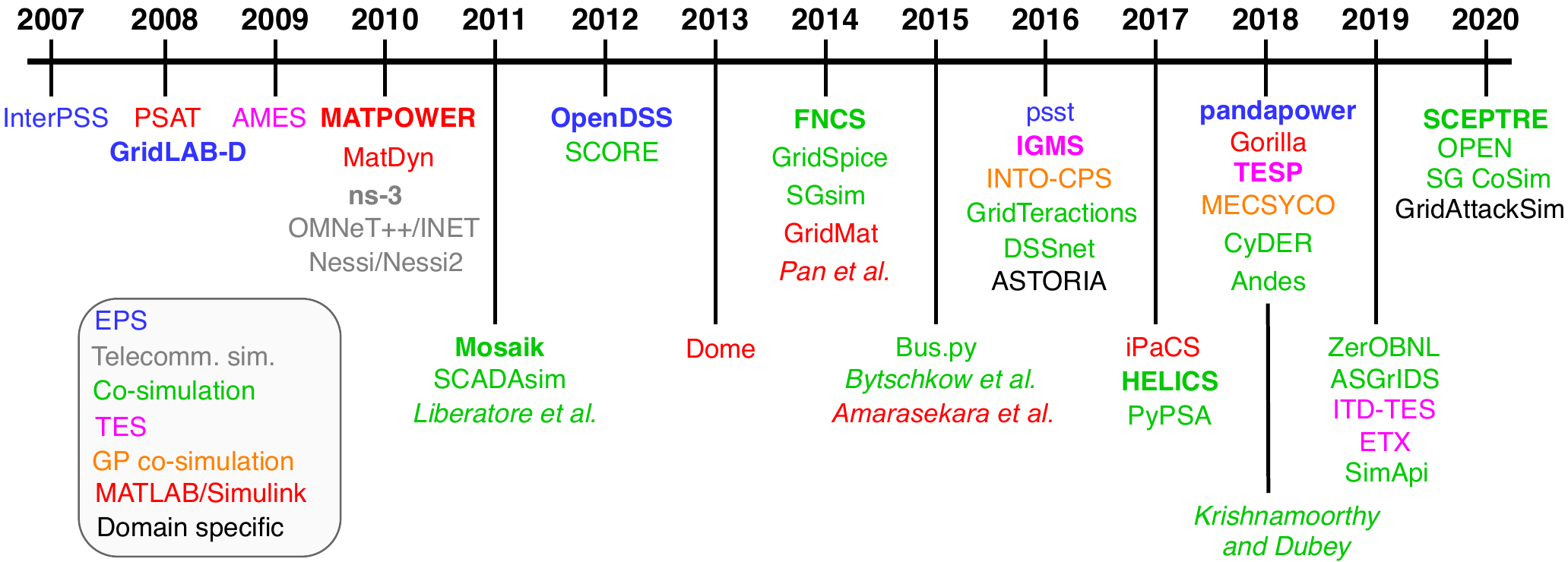}
	\caption{Non-comprehensive list of open-source/free tools and frameworks for co-simulating CPS.}
	\label{fig:tools}
\end{figure}

Figure~\ref{fig:tools} details the tools discussed here in chronological order.
We stress that there are more tools for co-simulating CPS, as this offers a non-comprehensive list.
The set of tools highlighted in bold face are tools that have sustained a high maturity degree over the years, with wide adoption in industry an academia.
The figure points out the sheer number of implementations modellers could choose to tackle CPS research.

Table~\ref{tab:frameworks-and-auxiliary} shows the frameworks and tools detailed here and their auxiliary tools.
One notices the recurrent use of MATLAB/Simulink, GridLAB-D, and ns-3 across several solutions.
\begin{table}[!htb]
\caption{Frameworks and auxiliary tools.}
\resizebox{\columnwidth}{!}{%
\centering
\begin{tabular}{|c|l||l|l|l|l|l|l|l|l|l|l|l|l|l|l|l|l|l|l|l|l|l|l|l|l|l|l|l|}
\hline
 \multicolumn{2}{|l|}{} & \STAB{\rotatebox[origin=c]{90}{20-sim, SysML}} & \STAB{\rotatebox[origin=c]{90}{BCVTB (EnergyPlus)}} & \STAB{\rotatebox[origin=c]{90}{EnergyPlus}} & \STAB{\rotatebox[origin=c]{90}{FESTIV}} & \STAB{\rotatebox[origin=c]{90}{Tk (GUI)}} & \STAB{\rotatebox[origin=c]{90}{FNCS/HELICS}} & \STAB{\rotatebox[origin=c]{90}{tc, netem}} & \STAB{\rotatebox[origin=c]{90}{GridDyn}} & \STAB{\rotatebox[origin=c]{90}{GridLAB-D}} & \STAB{\rotatebox[origin=c]{90}{~interface (tool agnostic)~}} & \STAB{\rotatebox[origin=c]{90}{LabVIEW}} & \STAB{\rotatebox[origin=c]{90}{MATLAB/Simulink}} & \STAB{\rotatebox[origin=c]{90}{MATPOWER}} & \STAB{\rotatebox[origin=c]{90}{MATTRANS}} & \STAB{\rotatebox[origin=c]{90}{Mininet}} & \STAB{\rotatebox[origin=c]{90}{Modelica}} & \STAB{\rotatebox[origin=c]{90}{Modelio Multi-Model}} & \STAB{\rotatebox[origin=c]{90}{Mosaik}} & \STAB{\rotatebox[origin=c]{90}{ns-2}} & \STAB{\rotatebox[origin=c]{90}{ns-3}} & \STAB{\rotatebox[origin=c]{90}{oct2pypower}} & \STAB{\rotatebox[origin=c]{90}{OMNeT++}} & \STAB{\rotatebox[origin=c]{90}{OpenDSS}} & \STAB{\rotatebox[origin=c]{90}{OpenDSS-G}} & \STAB{\rotatebox[origin=c]{90}{OpenModelica}} & \STAB{\rotatebox[origin=c]{90}{pandapower}} \\ \hline\hline
\multirow{3}{*}{\STAB{\rotatebox[origin=c]{90}{EPS}}} & Andes &  &  &  &  &  &  &  &  &  &  &  &  & \checkmark &  & \checkmark &  &  &  &  &  &  &  &  &  &  & \\ \cline{2-28}
& MATPOWER &  &  &  &  &  &  &  &  &  &  &  & \checkmark &  &  &  &  &  &  &  &  &  &  &  &  &  & \\ \cline{2-28}
& PYPOWER* &  &  &  &  &  &  &  &  &  &  &  &  & \checkmark &  &  &  &  &  &  &  & \checkmark &  &  &  &  & \\ \hline\hline
\multirow{19}{*}{\STAB{\rotatebox[origin=c]{90}{Co-simulation frameworks}}} & ASGriDS &  &  &  &  &  &  & \checkmark &  &  &  &  &  &  &  &  &  &  &  &  &  &  &  &  &  &  & \checkmark \\ \cline{2-28}
& Bus.py &  &  &  &  &  &  &  &  & \checkmark &  &  &  &  &  &  &  &  &  &  &  &  &  &  &  &  & \\ \cline{2-28}
& Bytschkow\cite{bytschkow2015combining} &  &  &  &  &  &  &  &  & \checkmark &  &  &  &  &  &  &  &  &  &  &  &  &  &  &  &  & \\ \cline{2-28}
& CyDER &  &  &  &  &  &  &  & \checkmark &  &  &  &  &  &  &  &  &  &  &  &  &  &  &  &  &  & \\ \cline{2-28}
& DSSnet &  &  &  &  &  &  &  &  &  &  &  &  &  &  & \checkmark &  &  &  &  &  &  &  & \checkmark &  &  & \\ \cline{2-28}
& FNCS/HELICS &  &  &  &  &  &  &  &  & \checkmark &  &  &  &  &  &  &  &  &  &  & \checkmark &  &  &  &  &  & \\ \cline{2-28}
& Gorilla &  &  &  &  &  &  &  &  &  & \checkmark &  & \checkmark &  &  &  &  &  &  &  &  &  &  &  &  &  & \\ \cline{2-28}
& GridSpice &  &  &  &  &  &  &  &  & \checkmark &  &  &  & \checkmark &  &  &  &  &  &  &  &  &  &  &  &  & \\ \cline{2-28}
& GridTeractions &  &  &  &  &  &  &  &  &  &  & \checkmark &  &  &  &  &  &  &  &  &  &  &  &  & \checkmark &  & \\ \cline{2-28}
& IGMS &  &  &  & \checkmark &  &  &  &  & \checkmark &  &  &  & \checkmark &  &  &  &  &  &  &  &  &  &  &  &  & \\ \cline{2-28}
& Krishnamoorthy\cite{krishnamoorthy2018framework} &  &  &  &  &  &  &  &  &  &  &  &  & \checkmark &  &  &  &  &  &  &  &  &  & \checkmark &  &  & \\ \cline{2-28}
& Liberatore\cite{liberatore2011smart} &  &  &  &  &  &  &  &  &  &  &  &  &  &  &  & \checkmark &  &  & \checkmark &  &  &  &  &  &  & \\ \cline{2-28}
& OPEN &  &  &  &  &  &  &  &  &  &  &  &  &  &  &  &  &  &  &  &  &  &  &  &  &  & \checkmark\\ \cline{2-28}
& SCADAsim &  &  &  &  &  &  &  &  &  &  &  & \checkmark &  &  &  &  &  &  &  &  &  & \checkmark &  &  &  & \\ \cline{2-28}
& SGsim &  &  &  &  &  &  &  &  &  &  &  &  &  &  &  &  &  &  &  &  &  & \checkmark & \checkmark &  &  & \\ \cline{2-28}
& SG Cosim &  &  &  &  &  &  &  &  &  &  &  &  &  &  &  &  &  & \checkmark &  & \checkmark &  &  & \checkmark &  &  & \\ \cline{2-28}
& SimApi &  & \checkmark &  &  &  &  &  &  &  &  &  &  &  &  &  &  &  &  &  &  &  &  &  &  &  & \\ \cline{2-28}
& ZerOBNL &  &  &  &  &  &  &  &  &  &  &  &  &  &  &  &  &  &  &  &  &  &  &  &  &  & \checkmark\\ \hline\hline
\multirow{6}{*}{\STAB{\rotatebox[origin=c]{90}{MATLAB}}} & Amarasekara\cite{amarasekara2015co} &  &  &  &  &  &  &  &  &  &  &  & \checkmark &  &  &  &  &  &  &  & \checkmark &  &  &  &  &  & \\ \cline{2-28}
& Dome &  &  &  &  &  &  &  &  &  &  &  & \checkmark &  &  &  &  &  &  &  &  &  &  &  &  &  & \\ \cline{2-28}
& GridMat &  &  &  &  &  &  &  &  &  &  &  & \checkmark &  &  &  &  &  &  &  &  &  &  &  &  &  & \\ \cline{2-28}
& iPaCS &  &  &  &  &  &  &  &  &  &  &  &  &  & \checkmark &  &  &  &  &  & \checkmark &  &  &  &  &  & \\ \cline{2-28}
& MatDyn &  &  &  &  &  &  &  &  &  &  &  & \checkmark &  &  &  &  &  &  &  &  &  &  &  &  &  & \\ \cline{2-28}
& Pan\cite{pan2016ns3} &  &  &  &  &  &  &  &  &  &  &  & \checkmark &  &  &  &  &  &  &  & \checkmark &  &  &  &  &  & \\ \hline\hline
\multirow{4}{*}{\STAB{\rotatebox[origin=c]{90}{TES}}} & AMES &  &  &  &  &  &  &  &  & \checkmark &  &  &  &  &  &  &  &  &  &  &  &  &  &  &  &  & \\ \cline{2-28}
& Hansen\cite{hansen2017evaluating} &  &  &  &  &  & \checkmark &  &  & \checkmark &  &  &  &  &  &  &  &  &  &  & \checkmark &  &  &  &  &  & \\ \cline{2-28}
& ITD-TES &  &  &  &  &  & \checkmark &  &  & \checkmark &  &  &  &  &  &  &  &  &  &  & \checkmark &  &  &  &  &  & \\ \cline{2-28}
& TESP &  &  & \checkmark &  &  & \checkmark &  &  & \checkmark &  &  &  &  &  &  &  &  &  &  & \checkmark &  &  &  &  &  & \\ \hline\hline
\STAB{\rotatebox[origin=c]{90}{G}} & Maestro/INTO-CPS & \checkmark &  &  &  &  &  &  &  &  &  &  &  &  &  &  &  & \checkmark &  &  &  &  &  &  &  & \checkmark &  \\ \hline\hline
\multirow{2}{*}{\STAB{\rotatebox[origin=c]{90}{D}}} & ASTORIA &  &  &  &  &  &  &  &  &  &  &  &  &  &  &  &  &  & \checkmark &  & \checkmark &  &  &  &  &  &  \\ \cline{2-28}
& GridAttackSim &  &  &  &  & \checkmark & \checkmark &  &  &  &  &  &  &  &  &  &  &  &  &  &  &  &  &  &  &  & \\ \hline\hline
\multicolumn{28}{|l|}{\textbf{Legend: }G: General Purpose; D: Domain Specific. *PYPOWER is \textit{based} on MATPOWER.} \\ \hline
\end{tabular}
}
\label{tab:frameworks-and-auxiliary}
\end{table}

We now focus our attention on the set of tools that are mature and stable, employed by researchers throughout the years for a wealth of models and propositions.
Amidst the plethora of frameworks and tools listed here some are highlighted over the years such as GridLAB-D, OpenDSS, pandapower, Mosaik, FNCS/HELICS, SCEPTRE, GridSpice, ZerOBNL, IGMS, TESP, MATPOWER, PyPSA, PYPOWER, BCVTB, EnergyPlus, OMNeT++/INET, and ns-3.
Because there are many features to consider in CPS infrastructures (market, transmission, distribution, transmission-distribution, telecommunications with multiple protocols/technologies, PMU/PDC, SCADA), it is difficult to compare solutions.

We now focus our attention on the set of tools that are mature and stable, employed by researchers throughout the years for a wealth of models and propositions.
Table~\ref{table:selected-tools} shows a list of stable tools (EPS, co-simulation, and TES) for modelling the SG.
The table shows the size of each tool, in Mega Bytes (MB), where the method used was to download the latest .zip file directly from the download site or GitHub link.

\begin{table}[t]
\caption{Selected tools for modelling the Smart Grid.}
\begin{center}
\begin{tabular}{|c|l|c|c|c|c|c|l|}\hline
\multicolumn{1}{|c|}{\multirow{4}{*}{\rotate{Type}}} & \multicolumn{1}{c|}{\multirow{3}{*}{Tool /}} & \multicolumn{1}{c|}{\multirow{4}{*}{\rotate{Size MB\,}}} & \multicolumn{1}{c}{\multirow{4}{*}{License}} & \multicolumn{1}{|c}{\multirow{3}{*}{Major}} & \multicolumn{1}{|c|}{\multirow{4}{*}{\rotate{Platform}}} & \multirow{3}{*}{Version} & \multicolumn{1}{c|}{\multirow{3}{*}{Toolchain/auxiliary tools}} \\
& \multicolumn{1}{c|}{\multirow{3}{*}{Framework}} & &  & \multirow{3}{*}{Language} & & \multirow{3}{*}{11/2020} & \multicolumn{1}{c|}{\multirow{3}{*}{Supporting frameworks}} \\ 
 & &  &  &  &  & &  \\ 
 & &  &  &  &  & &  \\ \hline\hline
\multirow{3}{*}{\rotate{EPS}} & GridLAB-D & 74 & BSD & C/C++ & Linux & 4.1.0 & Python, MATLAB \\ \cline{2-8}
 & pandapower & 27 & BSD & Python & Mult. & 2.4.0 & PYPOWER, pandas \\ \cline{2-8}
 & OpenDSS & 27 & BSD & Pascal & Mult. & 9.0.0 & OpenDSS-G (GUI) \\ \hline\hline
\multirow{3}{*}{\rotate{Co-Sim}} & FNCS & 0.7 & BSD & C/C++ & Linux & 2.3.2 & GridLAB-D, ns-3 \\ \cline{2-8}
 & OPEN & 75 & Apache2 & Python & Mult. & -- & pandapower, several dep. \\ \cline{2-8}
 & ZerOBNL & 1.1 & Apache2 & Python & Mult. & 1.1 & zeroMQ, Docker \\ \hline\hline
\rotate{TES} & TESP & 151 & BSD & C++ & Linux & 0.9.4 & 
\multicolumn{1}{l|}{
                    \begin{tabular}[c]{@{}l@{}}
                      GridLAB-D, FNCS/HELICS, \\
                      ns-3, EnergyPlus, PYPOWER\\
                    \end{tabular}} \\ \hline
\end{tabular}
\end{center}
\label{table:selected-tools}
\end{table}

The tools presented in the table use BSD or Apache2 licensing.
The sizes are quite reasonable apart from TESP (151 MB), requiring around 25 MB (pandapower, OpenDSS) or 75 MB (GridLAB-D, OPEN).
Co-simulation requires large system modelling and timely execution.
That is the main reason as to why high-performance languages are preferred such as C/C++ in GNU/Linux.
We point out that Python could also be used as a scripting language to fire executions and parse output, generating graphs for analysis.
Python has many advantages such as an extensive API and easy interface mechanisms to connect among solutions.
However, even with substantial code modification, depending on the model, Python falls short on performance.

A testbed mimics an infrastructure so researchers may conduct experiments without worrying about causing damages to the real systems in production.
Besides Deter, mentioned earlier, there is a considerable amount of work towards SCADA in terms of testbeds~\cite{hahn2013cyber,mallouhi2011testbed}.

As general comments, power simulators are sensitive when changing energy parameters as the steady-state computation may start yielding skewed results.
That certainly hinders analysis as incremental changes in models may not reflect reality, specially when the modeller is unfamiliar with the electrical engineering domain.

On a TES note, there is an increased interest in looking at ITD systems as they may impact dynamic pricing or related mechanisms in place.
GridSpice offered a solution that \textit{``blurred the boundaries between generation, transmission, distribution, and markets''} as stated by the authors.
This is an interesting concept as more abstractions across layers are needed for effective modelling of other SG behaviours.

Considered individually, the simulation engines are quite strong, however, problems start when they are integrated altogether.
For instance, GridLAB-D, pandapower, EnergyPlus, ns-3, MATPOWER, or OpenDSS are stable and mature tools.
When modellers and developers offer integration features, then we have observed that the novel solution does not share the same amount of stability and endurance.
There is also a lack of cross-platform validation to ensure that the simulations are yielding out appropriate results for deeper inspections.

\section{Conclusion}\label{s:conclusion}
As general remarks, there is research to conduct on increasing frameworks and models reproducibility.
To reach this objective one must create and share model repositories, datasets, logging data, or measurements.
Modellers choose the framework they are most familiar with, however, in terms of models, there is a lack of published artefacts so other researchers may extend, adapt, or augment existing functionalities.

Lack of cross-platform validation for frameworks is another disconcerting problem in this field.
There is a lot of models and frameworks but seldom validation efforts are presented for consistent scrutiny.
Sometimes the tool is available for download, however, the models and datasets are missing, or even in some cases configuration details (or options) are omitted, preventing proper execution.

Future research efforts will study models from different application domains to investigate primitives and parameters for tackling most impactful SG's challenges.
One direction is consider the use of Markovian approaches to model the SG and DER such as Arnaboldi et al. (2020)~\cite{arnaboldi2020modelling} who
modelled Load Changing Attacks over different mixes of power generation units.
Another is to use domain specific testbeds for addressing design shortcomings of SG as well as training stakeholders~\cite{ashok2019multi}.
The idea is to offer capabilities to detect and respond to abnormal situations in the SG as discussed by Henze et al. (2020) with a case study in cybersecurity~\cite{henze2020poster}.

\setlength{\LTleft}{0pt}

\section*{Acronyms}

\begin{longtable}{ll}
BES      & Building Energy Systems \\
CIM      & Common Information Model \\
COTS     & Commercial-Off-The-Shelf \\
CPES     & Cyber-Physical Energy Systems \\
CPF      & Continuation Power Flow \\
CPS      & Cyber-Physical System \\
DER      & Distributed Energy Resources \\
DEVS     & Discrete eVent system Specification \\
DoS      & Denial-of-Service \\
EMS      & Energy Management Systems \\
EPS      & Electric Power Simulator \\
EV       & Electric Vehicles \\
FDI      & False Data Injection \\
FMI      & Functional Mock-up Interface \\
FMU      & Functional Mock-up Units \\
HIL      & Hardware-in-the-Loop \\
ICS      & Industrial Control Systems \\
ICT      & Information and Communication Technologies \\
IDS      & Intrusion Detection Systems \\
ITD      & Integrated Transmission and Distribution \\
JADE     & Java Agent DEvelopment \\
MATLAB   & MATrix LABoratory \\
MTU      & Master Terminal Units \\
M\&S     & Modelling and Simulation \\
OMF      & Open Modelling Framework \\
OPF      & Optimal Power Flow \\
PDC      & Phasor Data Concentrators \\
PF       & Power Flow \\
PMU      & Phasor Management Units \\
PNNL     & Pacific Northwest National Laboratory \\
PSLF     & Positive Sequence Load Flow \\
PV       & Photo-Voltaic \\
RER      & Renewable Energy Resources \\
RTU      & Remote Terminal Units \\
SCADA    & Supervisory Control and Data Acquisition \\
SDN      & Software Defined Networks \\
SG       & Smart Grid \\
UC       & Unit Commitment \\
WTPCS    & Wind Turbines Power Converter Systems \\
\end{longtable}

\section*{Acronyms for tools}
\begin{longtable}{ll}
AMES     & Agent-based Modeling of Electricity Systems \\
ASGriDS  & Asynchronous Smart Grid Simulation \\
ASTORIA  & Attack Simulation Toolset for Smart Grid Infrastructures \\
BCVTB    & Building Controls Virtual Test Bed \\
CyDER    & Cyber physical co-simulation platform for Distributed Energy Resources \\
Deter    & cyber-DEfense Technology Experimental Research \\
DSSnet   & Distribution System Solver Network \\
EPOCHS   & Electric Power and cOmmunication synCHronizing Simulator \\
FESTIV   & Flexible Energy Scheduling Tool for Integration of Variable Generation \\
FNCS     & Framework for Network Co-Simulation \\
GECO     & Global Event-driven CO-simulation framework \\
HELICS   & Hierarchical Engine for Large-scale Infrastructure Co-Simulation \\
IGMS     & Integrated Grid Modeling System \\
INSPIRE  & INtegrated co-Simulation of Power and ICT systems for Real-time Evaluation \\
InterPSS & Internet-technology Based Power System Simulator \\
INTO-CPS & Integrated Tool Chain for Model-based Design of CPS \\
iPaCS    & integrative Power and Cyber Systems \\
JADE     & Java Agent DEvelopment \\
MATTRANS & MATLAB/Simulink Power System Transient Stability Simulation \\
MECSYCO  & Multi-agent Environment for Complex SYstems CO-simulation \\
Nessi    & Network Security Simulator \\
ns-3     & Network Simulator version 3 \\
OMF      & Open Modelling Framework \\
OMNeT++  & Objective Modular Network Testbed in C++ \\
OpenDSS  & Open Distribution System Simulator \\
OPEN     & Open Platform for Energy Networks \\
PSAT     & Power System Analysis Toolbox \\
PSCAD    & Power Systems Computer Aided Design \\
PSS      & Power System Simulator \\
psst     & Power System Simulation Toolbox \\
PyPSA    & Python for Power System Analysis \\
RTDS     & Real-Time Digital Simulator \\
SCORE    & Smart-Grid Common Open Research Emulator \\
SEPSS    & Scalable Electric Power System Simulator \\
SG-Cosim & SmartGrid Cosimulation \\
SOSPO-SP & Secure Operation of Sustainable Power Systems Simulation Platform \\
TASSCS   & Testbed for Analyzing Security of SCADA Control Systems \\
TES      & Transactive Energy Systems \\
TESP     & Transactive Energy Simulation Platform \\
VPST     & Virtual Power System Testbed \\
ZerOBNL  & Zero OBvious Node Link co-simulator \\
\end{longtable}

\medskip

\printbibliography

\end{document}